\documentclass[a4paper]{article}

\usepackage{INTERSPEECH2022}
\usepackage{graphicx}
\usepackage{multicol}
\usepackage{multirow}
\usepackage{subcaption}
\usepackage{comment}
\usepackage{xcolor}
\usepackage{float}
\def\wavtovecu/{wav2vec-U}
\def\wavtovectwo/{wav2vec 2.0}
\usepackage{hyperref}

\title{Unsupervised Text-to-Speech Synthesis by Unsupervised Automatic Speech Recognition}
\name{Junrui Ni$^{1^{*}}$, Liming Wang$^{1^{*}}$, Heting Gao$^{1^{*}}$, Kaizhi Qian$^2$, Yang Zhang$^2$, Shiyu Chang$^3$, Mark Hasegawa-Johnson$^1$}
\address{
  $^1$University of Illinois at Urbana-Champaign\\
  $^2$MIT-IBM Watson AI Lab
  $^3$University of California, Santa Barbara}
\email{\{junruin2,lwang114,hgao17,jhasegaw\}@illinois.edu, \{kqian, yang.zhang2\}@ibm.com, chang87@ucsb.edu}

\begin{document}

\maketitle
\def\thefootnote{*}\footnotetext{These authors contributed equally to this work}\def\thefootnote{\arabic{footnote}}
\begin{abstract}
  An unsupervised text-to-speech synthesis (TTS) system learns to generate speech waveforms corresponding to any written sentence in a language by observing: 1) a collection of untranscribed speech waveforms in that language; 2) a collection of texts written in that language without access to any transcribed speech. Developing such a system can significantly improve the availability of speech technology to languages without a large amount of parallel speech and text data. This paper proposes an unsupervised TTS system based on an alignment module that outputs pseudo-text and another synthesis module that uses pseudo-text for training and real text for inference. Our unsupervised system can achieve comparable performance to the supervised system in seven languages with about 10-20 hours of speech each. A careful study on the effect of text units and vocoders has also been conducted to better understand what factors may affect unsupervised TTS performance. The samples generated by our models can be found at \url{https://cactuswiththoughts.github.io/UnsupTTS-Demo}, and our code can be found at \url{https://github.com/lwang114/UnsupTTS}.
\end{abstract}

\section{Introduction}
Text-to-speech (TTS) synthesis is an essential component of a spoken dialogue system. While capable of generating high-fidelity, human-like speech for languages such as English and Mandarin, the existing state-of-the-art TTS systems such as Tacotron 1\&2~\cite{Wang2017-tacotron,Shen2018-tacotron2}, Deep Voice 3~\cite{Ping2018-deep-voice3}, FastSpeech~\cite{Ren2019-fastspeech} and Transformer TTS~\cite{Li2019} are trained with a large amount of parallel speech and textual data. The reliance on a large amount of transcribed speech makes such systems impractical for the majority of the languages in the world. Training a supervised text-to-speech (TTS) system requires dozens of hours of single-speaker high-quality recordings \cite{Xu20-lrspeech}, but collecting a large amount of single-speaker, clean, and transcribed speech corpus can be quite time-consuming and expensive~\cite{Park2019-css10}. A potential way to relax such a requirement is to use \emph{non-parallel} untranscribed speech and text corpora in the same language. Such corpora are much easier to obtain in practice since no human annotators are required in the data collection process, thanks to the abundance of text data on the Internet. Learning to perform TTS using non-parallel speech and text, or \emph{unsupervised TTS}, poses unique challenges: first, standard supervised training criteria such as autoregressive mean-squared error are no longer applicable; further, to learn a latent alignment between the spoken frames and text units, the model now needs to search over every utterance and every transcript in the entire corpus instead of limiting the search space within a single utterance-transcript pair. 

This paper proposes the first model for solving the unsupervised TTS problem. We decompose training the model into two tasks, learning an alignment module that assigns a single pseudo-transcript to each utterance and learning a synthesis module that learns from pseudo-text and utterance pairs. The alignment module is motivated by the best publicly available unsupervised ASR system, \wavtovecu/~\cite{Baevski2021-wav2vec-u}, and can generalize seamlessly to future updates of the \wavtovecu/ model. We conduct our unsupervised TTS experiments on seven languages. We further provide an in-depth analysis of the effect of several components on unsupervised TTS performance, including the grapheme-to-phoneme (G2P) converter and the vocoder.

\begin{figure*}
    \centering
    \includegraphics[width=0.75\textwidth]{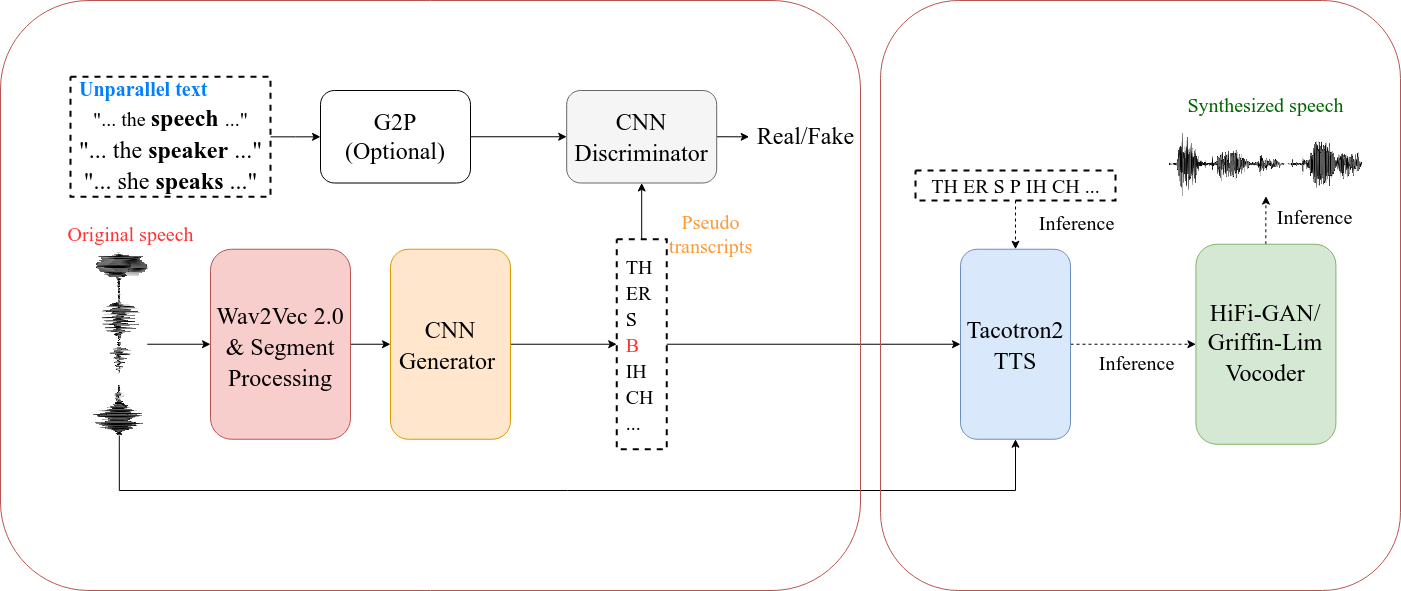}
    \caption{Network architecture for unsupervised speech synthesis, splitted into an alignment module (left) and a synthesis module (right)}
    \label{fig:unsup_tts}
\end{figure*}

\section{Related works}
Several recent works have attempted to develop TTS systems for low-resource scenarios. One direction of research is to replace ground truth phoneme or grapheme labels required for supervised TTS with other units obtained with less or no supervision, such as articulatory features~\cite{Muthukumar2014}, or acoustic units discovered by self-supervised speech representation models such as vector-quantized variational auto-encoder (VQ-VAE)~\cite{Liu20-seqae, Zhang20-unsupttslr} and HuBERT~\cite{Lakhotia2021,Polyak2021,Kharitonov2021}.
The unsupervised, textless approach can be applied to any language, including those without any written form. However, the performance of such a system is limited by the quality of the acoustic units used, which can be quite noisy due to the difficulty of acoustic unit discovery. To address this limitation, ~\cite{Hasegawa-Johnson2017,Wang2020-show-and-speak,Hsu-etal-2021,Effendi2021} have studied the use of other sensory modalities such as images in place of textual transcripts as a weaker form of supervision for conditional generation of speech, or ``TTS without T'', using various attention mechanisms over the visual features. Another approach to address this issue is to allow a small amount of transcribed speech and train the TTS in a semi-supervised fashion \cite{Ren2019-almost-unsup-tts, Xu20-lrspeech}. Specifically, \cite{Ren2019-almost-unsup-tts} leveraged unpaired speech and text data by constructing pseudo-corpora via dual transformation between ASR and TTS systems with on-the-fly refinement followed by knowledge distillation, while LRSpeech~\cite{Xu20-lrspeech} trained an ASR and a TTS system that used only several minutes of paired single-speaker, high-quality speech for TTS, and several hours of low-quality, multi-speaker data for ASR.

Our approach is motivated by the most recently published unsupervised automatic speech recognition (ASR) system~\cite{Baevski2021-wav2vec-u}, which learns to recognize phones by leveraging pre-trained speech representations and an unpaired text corpus. Other works on unsupervised ASR typically try to match the empirical prior and posterior distributions of phonemes either using cross-entropy~\cite{Yeh2019-unsup-asr} or adversarial loss~\cite{Chen2019}. Using powerful self-supervised, pre-trained acoustic features such as \wavtovectwo/~\cite{Baevski2020-wav2vec2} and a generative adversarial network (GAN)-based system, the adversarial approach with self-training achieves comparable performance to its supervised counterpart on large-scale speech datasets for multiple languages~\cite{Baevski2021-wav2vec-u}. 

\section{Proposed method}
The proposed unsupervised TTS system contains two modules: an alignment module to obtain pseudo-text for each utterance and a synthesis module that trains on pseudo-text. We evaluate the proposed unsupervised TTS system in English, as well as in six other languages (Hungarian, Spanish, Finnish, German, and Japanese).

\subsection{Alignment Module}
Motivated by \wavtovecu/~\cite{Baevski2021-wav2vec-u}, our alignment module greedily proposes a pairing relationship (which we denote as an alignment) between real speech utterances and pseudo-transcripts. 
The alignment module follows a two-step approach: GAN training and self-training. In the GAN training step, a 1-layer CNN acts as the generator, which takes the segment representations extracted from a pre-trained \wavtovectwo/ model \cite{Baevski2020-wav2vec2} and outputs a sequence of distributions over text units, where consecutive segments with the same $\operatorname{argmax}$ value are collapsed. The discriminator, a 3-layer CNN trained against the generator, tries to tell which source (real or generated) the input sequence is from. This is achieved by iteratively maximizing the likelihood of the generated phoneme sequence to train the generator and minimizing the binary cross-entropy loss to train
the discriminator.
In addition, since GAN training can be very unstable, we search over the weights for regularization losses such as gradient penalty loss, segment smoothness penalty, and phoneme diversity loss as described in \cite{Baevski2021-wav2vec-u}. We also validate the model with 50-100 transcribed utterances from the corpus to ensure convergence instead of using the unsupervised metric in \cite{Baevski2021-wav2vec-u}. After GAN training, greedy decoding is applied to the generator's output over the training set. We then train a hidden Markov model (HMM) with framewise speech representations extracted from a \wavtovectwo/ model as input and pseudo-text decoded by the generator as output. Finally, we decode the entire corpus again using the newly-trained HMM to obtain pseudo-transcripts for the supervised TTS system. Except for English, we opt not to further fine-tune a \wavtovectwo/ model with the pseudo-transcripts from the HMM.

\subsection{Synthesis Module}
The synthesis module uses the proposed alignment pairs from the alignment module and learns to synthesize speech from pseudo-text.
Our synthesis module is motivated by Tacotron 2~\cite{Shen2018-tacotron2} with guided attention loss~\cite{Tachibana2018-DCTTS}. During training, we perform an unsupervised model selection process by feeding the module with pseudo-transcripts when computing validation loss. During evaluation, ground truth transcripts are used as inputs to the synthesis module. Character error rates (CER) and word error rates (WER) are used to measure how much linguistic content is preserved by the synthesis module.
We train a fully supervised TTS system using real text instead of pseudo-text and calculate the CER and WER on the same subset for comparison. To obtain the CER and the WER on each language, we either directly use a publicly available \wavtovectwo/ speech recognizer (for English) or fine-tune a pre-trained \wavtovectwo/ model on each language individually.

\section{Experiments}
\subsection{Unsupervised TTS on English}\label{sec:tts_english}
We first evaluated the two-stage unsupervised TTS system on English. To train the alignment module, we used speech utterances from the 24-hour single-speaker LJSpeech corpus \cite{ljspeech17} and text samples from the LibriSpeech language modeling corpus \cite{Panayotov15-LibriSpeech}.
We set aside about 300 utterances for validation and about 500 utterances for testing. We kept the ground truth transcripts for validation and test sets and used the rest for training without ground truth transcripts. The speech representations were extracted using a publicly available \wavtovectwo/ Large model trained on LibriLight \cite{librilight}, and the segment representations were built following the pre-processing procedures in \cite{Baevski2021-wav2vec-u}. 

\begin{table}[ht]
    \caption{Alignment module results on the LJSpeech dataset using English \wavtovectwo/ pre-trained features}
    \label{tab:unsup_asr_w2v_lj}
    \centering
    \begin{tabular}{lccccccc}
    \toprule
    \multirow{2}{*}{Language} & \multirow{2}{*}{Duration (hr)} & \multicolumn{2}{c}{Unsup ASR (PER)} \\
    \cmidrule(lr){3-4}
    & & No ST & ST \\
    \midrule
    \midrule
    English & 24 & 12.37 & 3.59 \\
    \bottomrule
    \end{tabular}
\end{table}

\begin{table}[ht]
    \caption{Unsupervised TTS results on the LJSpeech dataset using English \wavtovectwo/ pre-trained features}
    \label{tab:unsup_tts_w2v_lj}
    \centering
    \begin{tabular}{lcccc}
        \toprule
         Language & \multicolumn{2}{c}{Unsup TTS} & \multicolumn{2}{c}{Supervised TTS}\\
         \cmidrule(lr){2-3}\cmidrule(lr){4-5}
         & CER & WER & CER & WER\\
        \midrule
        \midrule
        English & 4.56 & 11.95 & 3.93 & 10.76 \\
        \bottomrule
    \end{tabular}
\end{table}

The non-parallel text samples used for training, as well as the ground truth transcripts for the validation and test utterances, were converted to phones using a grapheme-to-phoneme (G2P) converter \cite{g2pE2019}. The best weights for the auxiliary penalties of the GAN system, i.e., code penalty, gradient penalty, and smoothness weight, c.f.~\cite{Baevski2021-wav2vec-u}, were determined by grid search, and we chose the best model based on its PER over the validation set after 150k steps with a batch size of 160. GAN training is sometimes unstable in ways that we could only detect by using 50-100 supervised validation examples, which were the only places during training where we used paired data. The results of this stage is shown in Table~\ref{tab:unsup_asr_w2v_lj}. After determining the best GAN model, its output phone sequence was then refined using a self-training (ST) process \cite{Baevski2021-wav2vec-u} as follows. First, we used framewise \wavtovectwo/ features after PCA transformation as input and pseudo phone sequences transcribed by the generator as targets to train a triphone HMM. The triphone output from the HMM was decoded into words with an HCLG decoding graph, and we further fine-tuned a \wavtovectwo/ Large model using the pseudo character targets obtained from the above step, under the Connectionist Temporal Classification (CTC) loss \cite{Graves06-CTC}. Both steps were validated with the corresponding pseudo-text for the validation set. As shown in Table~\ref{tab:unsup_asr_w2v_lj}, ST reduces the phone error rate on the test set by 70\% relative and provides very accurate transcripts for the second-stage TTS system. We used the publicly available  \wavtovecu/ model in the Fairseq toolkit \cite{Ott19-fairseq} to train the GAN and used the Kaldi toolkit \cite{Povey11-kaldi} to train the triphone HMM and to build the decoding graph.

\begin{figure}
    \centering
    \includegraphics[width=\linewidth,trim={0.62cm 0cm 0.55cm 0.2cm},clip]{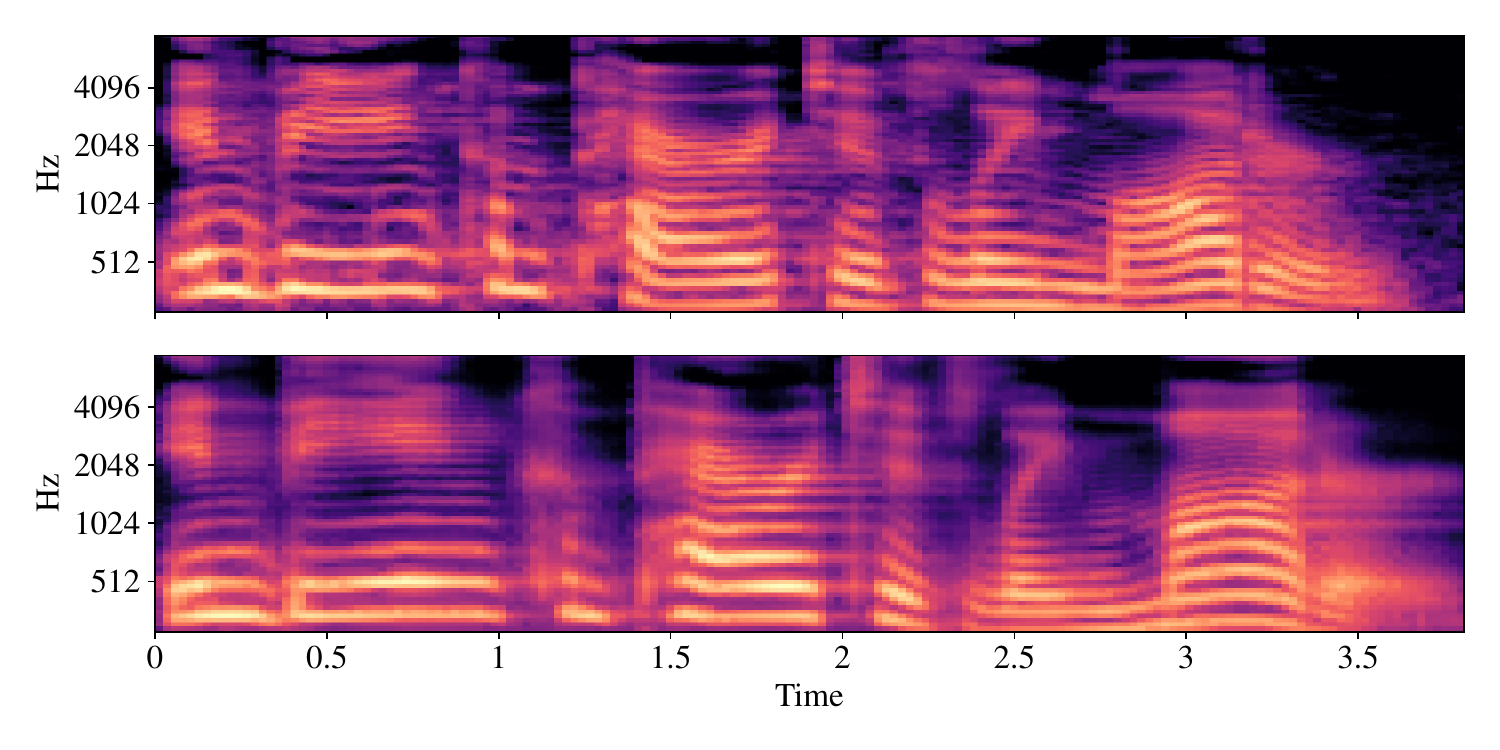}
    \caption{Mel-spectrograms for ground truth (upper) and synthetic speech by the unsupervised TTS model (lower) for the English sentence ``in being comparatively modern.''}
    \label{fig:visualize_sgram}
\end{figure}

To train the synthesis module, we used the Tacotron 2 \cite{Shen2018-tacotron2} implementation in ESPnet~\cite{Hayashi20-espnet}. The ESPnet implementation follows the original Tacotron 2 model, except that another guided attention loss \cite{Tachibana2018-DCTTS} was calculated on top of the encoder-decoder attention matrix so that it is not too far from being diagonal. During training, the synthesis module takes pseudo phone transcripts as inputs, and outputs 80-dimensional mel-spectrograms. These pseudo phone transcripts are converted by G2P from the word-level hypotheses generated by the fine-tuned \wavtovectwo/ model (in the final step of alignment module training). The synthesis module was trained for 80 epochs, with the same validation and test splits used for training the alignment module. During validation of the synthesis module, we calculated the reconstruction loss based on pseudo-text instead of real text.
During testing, we fed the trained synthesis module with real, phonemicized text transcripts for the test set to obtain mel-spectrograms and synthesized raw audios with HiFi-GAN \cite{Kong20-hifigan}. We calculated the CERs and raw WERs without additional language models using a publicly available \wavtovectwo/ Large model fine-tuned on LibriSpeech. Table~\ref{tab:unsup_tts_w2v_lj} shows the two error rates on the synthesized test utterances using our proposed unsupervised system (Unsup TTS). Compared with a fully-supervised Tacotron 2 model trained and validated with real, phonemicized text transcripts, our unsupervised system only lags behind 0.63\% absolute in terms of CER and 1.19\% absolute in terms of WER. Figure~\ref{fig:visualize_sgram} plots the mel-spectrogram of a synthetic speech example by our unsupervised model, which shows that except for the temporal patterns, the mel-spectrogram by the unsupervised TTS looks very similar to the ground truth with very little loss of linguistic content.

\subsection{Unsupervised TTS on CSS10 Languages}
\label{sec:tts_css10}
\begin{table}[ht]
    \caption{Unsupervised ASR results on the CSS10 dataset using English \wavtovectwo/ pretrained features}
    \label{tab:unsup_asr_w2v_css}
    \centering
    \begin{tabular}{lccccccc}
    \toprule
    \multirow{2}{*}{Language} & \multirow{2}{*}{Duration (hr)} & \multicolumn{2}{c}{Unsup ASR (CER)} \\
    \cmidrule(lr){3-4}
     & & No ST & ST \\
    \midrule
    \midrule
    Japanese  & 15 & 26.12 & 17.81 \\
    Hungarian & 10 & 25.08 & 15.26 \\
    Spanish   & 24 & 20.80 & 14.57 \\
    Finnish   & 10 & 29.78 & 21.00 \\
    German    & 17 & 26.31 & 19.47 \\
    Dutch & 14 & 45.65 & 39.24 \\
    %
    \bottomrule
    \end{tabular}
\end{table}

\begin{table}[ht]
    \caption{Unsupervised TTS results on the CSS10 dataset using English \wavtovectwo/ pretrained features}
    \label{tab:unsup_tts_w2v_css}
    \centering
    \begin{tabular}{lcccc}
    \toprule
    \multirow{2}{*}{Language} & \multicolumn{2}{c}{Unsup TTS} & \multicolumn{2}{c}{Supervised TTS}\\
    \cmidrule(lr){2-3}\cmidrule(lr){4-5}
     & CER & WER & CER & WER\\
    \midrule
    \midrule
    Japanese  & 17.98 & 47.81 & 17.87 & 36.23\\
    Hungarian & 27.78 & 76.82 & 18.05 & 63.14 \\
    Spanish   & 23.03 & 55.52 & 18.19 & 36.74 \\
    Finnish   & 36.05 & 84.46 & 22.84 & 58.67 \\
    German    & 17.25 & 56.78 & 11.28 & 40.94 \\
    Dutch     & 53.01 & 89.41 & 34.53 & 76.71\\
    %
    \bottomrule
    \end{tabular}
\end{table}

\begin{table}[ht]
    \caption{The effect of different pretrained vocoders (Griffin-Lim, HiFi-GAN) on unsupervised TTS results for LJSpeech and various languages from CSS10}
    \label{tab:effect_speech_features}
    \centering
    \begin{tabular}{lcccc}
    \toprule
    \multirow{2}{*}{Language} & \multicolumn{2}{c}{Griffin-Lim} & \multicolumn{2}{c}{HiFi-GAN}\\
    \cmidrule(lr){2-3}\cmidrule(lr){4-5}
     & CER & WER & CER & WER\\
    \midrule
    \midrule
    English & 5.02 & 12.83 & \textbf{4.56} & \textbf{11.95} \\
   Japanese & \textbf{17.98} & \textbf{47.81} & 20.58 & 54.09 \\
   Hungarian & 27.78 & 76.82 & \textbf{26.92} & \textbf{76.60}\\
   Spanish & \bf23.03 & \bf55.52 & 29.41 & 68.82  \\
    Finnish & \bf36.05 & \bf84.46 & 37.66 & 87.48   \\ 
    German & \textbf{17.25} & \textbf{56.78} & 18.45  & 59.90 \\
    \bottomrule
    \end{tabular}
\end{table}

\begin{table}[ht]
    \caption{The effect of different text units on unsupervised TTS using Griffin-Lim vocoder}
    \label{tab:unsup_tts_phoneme_vs_grapheme}
    \centering
    \begin{tabular}{lcccc}
    \toprule
    \multirow{2}{*}{Language}  & \multicolumn{2}{c}{Phoneme} & \multicolumn{2}{c}{Grapheme}\\
    \cmidrule(lr){2-3}\cmidrule(lr){4-5}
     & CER & WER & CER & WER\\
    \midrule
    \midrule
    Hungarian & \textbf{22.73} & \textbf{68.80} & 27.78 & 76.82 \\
    Finnish & \textbf{27.58} & \textbf{67.87} & 36.05 & 84.46  \\
    Dutch & \textbf{22.04} & \textbf{56.85} & 53.01 & 89.41 \\
    \bottomrule
    \end{tabular}
\end{table}
We evaluated our unsupervised TTS system on six additional languages: Japanese, Hungarian, Spanish, Finnish, German and Dutch from the CSS10 dataset~\cite{park2019css10}. The total duration of each language is listed in Table~\ref{tab:unsup_asr_w2v_css}. The experiments followed similar steps as the English experiment in Sec \ref{sec:tts_english}. The alignment module extracts speech representations from the same English \wavtovectwo/ Large model, followed by GAN training and self-training. The results for the alignment module are shown in Table~\ref{tab:unsup_asr_w2v_css}. There were still a few differences in details in this multilingual experiment. Due to resource limits, these multilingual experiments used a potentially easier setting where both the audio and text were drawn from the same CSS10 dataset with their paired relationship broken up, instead of from different datasets as for English. We did not convert graphemes into phonemes and directly used the characters in each language as text unit. We split the audio and text data into training and validation sets with a ratio of 99 to 1, leaving about 50 to 100 validation utterances depending on the dataset size. The self-training step of the first stage only contained a character-based HMM (instead of a triphone HMM with HCLG decoding) for generating pseudo-text, and we did not have a second step of fine-tuning a \wavtovectwo/ model as for English. During the evaluation of the synthesis module, we used a Griffin-Lim vocoder to synthesize the audios from the generated mel-spectrogram, and the results reported in Table~\ref{tab:unsup_tts_w2v_css} were calculated using audios from the Griffin-Lim vocoder instead of the HiFi-GAN vocoder. We switched to the Griffin-Lim vocoder because we empirically found that it yielded lower error rates on these languages. To calculate CER and raw WER, we fine-tuned a publicly available \wavtovectwo/ Base model for each language individually, using paired speech and character-level transcripts from each CSS10 corpus.

The multilingual results in Table~\ref{tab:unsup_tts_w2v_css} confirm the conclusions we reach in the English experiments. Although the self-training step is simplified to only a character-based HMM, the self-training step still greatly reduces the error rates by 25\% to 40\% relative to all the languages. Compared to the fully-supervised Tacotron 2 models trained using real text transcripts, the CERs of our unsupervised systems differ by only about 9\% absolute on average while requiring only a few paired utterances during validation. Further, we observe that the gap in WER between supervised and unsupervised TTS systems generally is about 10-20\% absolute for all languages except Finnish, a much larger gap than CER. We hypothesize that it may be due to the lack of a robust language model in the TTS systems, making it harder for the model to preserve word-level information when training with noisy (pseudo-)transcripts. Last but not least, we observe that the performance of the alignment module does not always limit the performance of unsupervised TTS. In the case of German, the TTS trained with pseudo-transcripts achieves a \emph{lower} CER compared to the alignment module alone, which suggests that the TTS has some internal mechanism to correct the noise in the pseudo-transcripts. 

\subsection{Comparison Between Griffin-Lim and HiFi-GAN}
\label{subsec:gl_vs_hifi}
A comparison between the error rates of using  Griffin-Lim and HiFi-GAN vocoders is presented in Table~\ref{tab:effect_speech_features}. We observe that the Griffin-Lim vocoder yields lower CERs and WERs than the HiFi-GAN vocoder in all languages except English and Hungarian, even though informal listening suggests that HiFi-GAN generates more natural speech with fewer artifacts. We hypothesize that HiFi-GAN works better for English because it is pre-trained on the English LJSpeech dataset and may not generalize very well when applied to datasets of different languages.

\subsection{Comparison Between Phoneme and Grapheme}
\label{subsec:ph_vs_gr}
We trained additional phoneme-based unsupervised TTS models in Hungarian, Finnish, and Dutch to study how the text units affect system performance. The training procedure was the same as that described in Sec~\ref{sec:tts_css10}, except that for training the alignment module, we converted the language-specific graphemes to the phonetic annotations, i.e., the International Phonetic Alphabet, using LanguageNet  G2Ps~\cite{hasegawa2020grapheme}. We then use the phonetic outputs from the HMM within the alignment module to train the synthesis module. The CERs and WERs are reported in Table~\ref{tab:unsup_tts_phoneme_vs_grapheme}. The table shows that the phone-based systems yield significantly lower error rates than the grapheme systems. As graphemes are the smallest functional unit of a writing system, it involves extra complexity on top of the phone systems. Thus, modeling the grapheme systems is harder than modeling the phone systems, as indicated by its higher error rates. The gap between grapheme and phoneme systems is considerably smaller for Hungarian and Finnish than for Dutch. One probable explanation is that spelling and phonetic transcription is far more regular for the former two languages than for Dutch.

\section{Conclusions}
In this work, we combined an alignment module and a synthesis module to build a unsupervised TTS system that trains without paired data. 
The final unsupervised TTS system demonstrates competitive intelligibility in English and a slight degradation in intelligibility in six other languages on the level of supervised TTS models. We further show that phonemes work better than graphemes as text units for our systems. In the future, we would like to explore unsupervised TTS with truly non-parallel datasets for languages other than English and ways to improve the stability for the alignment module.

\section{Acknowledgements}
This work is supported by IBM-UIUC Center for Cognitive Computing Systems Research (C3SR). We would like to thank one anonymous reviewer for insights on Sec \ref{subsec:ph_vs_gr}.

\bibliographystyle{IEEEtran}

\bibliography{reference}

\begin{thebibliography}{10}
\providecommand{\url}[1]{#1}
\csname url@samestyle\endcsname
\providecommand{\newblock}{\relax}
\providecommand{\bibinfo}[2]{#2}
\providecommand{\BIBentrySTDinterwordspacing}{\spaceskip=0pt\relax}
\providecommand{\BIBentryALTinterwordstretchfactor}{4}
\providecommand{\BIBentryALTinterwordspacing}{\spaceskip=\fontdimen2\font plus
\BIBentryALTinterwordstretchfactor\fontdimen3\font minus
  \fontdimen4\font\relax}
\providecommand{\BIBforeignlanguage}[2]{{%
\expandafter\ifx\csname l@#1\endcsname\relax
\typeout{** WARNING: IEEEtran.bst: No hyphenation pattern has been}%
\typeout{** loaded for the language `#1'. Using the pattern for}%
\typeout{** the default language instead.}%
\else
\language=\csname l@#1\endcsname
\fi
#2}}
\providecommand{\BIBdecl}{\relax}
\BIBdecl

\bibitem{Wang2017-tacotron}
\BIBentryALTinterwordspacing
Y.~Wang, D.~S. RJ~Skerry-Ryan, Y.~Wu, R.~J. Weiss, N.~Jaitly, Z.~Yang, Y.~Xiao,
  Z.~Chen, S.~Bengio \emph{et~al.}, ``Tacotron: Towards end-to-end speech
  synthesis,'' in \emph{arXiv}, 2017. [Online]. Available: \url{preprint
  arXiv:1703.10135}
\BIBentrySTDinterwordspacing

\bibitem{Shen2018-tacotron2}
J.~Shen, R.~Pang, R.~J. Weiss, M.~Schuster, N.~Jaitly, Z.~Yang, Z.~Chen,
  Y.~Zhang, Y.~Wang, R.~Skerry-Ryan, R.~A. Saurous, Y.~Agiomyrgiannakis, and
  Y.~Wu1, ``Natural tts synthesis by conditioning wavenet on mel spectrogram
  predictions,'' in \emph{ICASSP}, 2018.

\bibitem{Ping2018-deep-voice3}
W.~Ping, K.~Peng, A.~Gibiansky, S.~O. Arik, A.~Kannan, S.~Narang, J.~Raiman,
  and J.~Miller, ``Deep voice 3: 2000-speaker neural text-to-speech,'' in
  \emph{ICLR}, 2018.

\bibitem{Ren2019-fastspeech}
Y.~Ren, Y.~Ruan, X.~Tan, T.~Qin, S.~Zhao, Z.~Zhao, and T.-Y. Liu, ``Fastspeech:
  Fast, robust and controllable text to speech,'' in \emph{Advances in Neural
  Information Processing Systems}, 2019.

\bibitem{Li2019}
N.Li, S.Liu, Y.Liu, S.Zhao, and M.Liu, ``Neural speech synthesis with
  transformer network,'' in \emph{AAAI}, vol.~33, 2019, p. 6706–6713.

\bibitem{Xu20-lrspeech}
J.~Xu, X.~Tan, Y.~Ren, T.~Qin, J.~Li, S.~Zhao, and T.~Liu, ``{LRS}peech:
  Extremely low-resource speech synthesis and recognition,'' in \emph{KDD},
  2020, pp. 2802--2812.

\bibitem{Park2019-css10}
K.~Park and T.~Mulc, ``{CSS10}: A collection of single speaker speech datasets
  for 10 languages,'' \emph{Interspeech}, 2019.

\bibitem{Baevski2021-wav2vec-u}
A.~Baevski, W.-N. Hsu, A.~Conneau, and M.~Auli, ``Unsupervised speech
  recognition,'' in \emph{Neural Information Processing Systems}, 2021.

\bibitem{Muthukumar2014}
P.~K. Muthukumar and A.~W. Black, ``Automatic discovery of a phonetic inventory
  for unwritten languages for statistical speech synthesis,'' in \emph{ICASSP},
  2014, pp. 2594--2598.

\bibitem{Liu20-seqae}
A.~H. Liu, T.~Tu, H.~Lee, and L.~Lee, ``Towards unsupervised speech recognition
  and synthesis with quantized speech representation learning,'' in
  \emph{ICASSP}, 2020, pp. 7259--7263.

\bibitem{Zhang20-unsupttslr}
H.~Zhang and Y.~Lin, ``Unsupervised learning for sequence-to-sequence
  text-to-speech for low-resource languages,'' in \emph{Interspeech}, 2020, pp.
  3161--3165.

\bibitem{Lakhotia2021}
\BIBentryALTinterwordspacing
K.~Lakhotia, E.~Kharitonov, W.-N. Hsu, Y.~Adi, A.~Polyak, B.~Bolte, T.-A.
  Nguyen, J.~Copet, A.~Baevski, A.~Mohamed, and E.~Dupoux, ``On generative
  spoken language modeling from raw audio,'' in \emph{arXiv}, 2021. [Online].
  Available: \url{https://arxiv.org/pdf/2102.01192.pdf}
\BIBentrySTDinterwordspacing

\bibitem{Polyak2021}
\BIBentryALTinterwordspacing
A.~Polyak, Y.~Adi, J.~Copet, E.~Kharitonov, K.~Lakhotia, W.-N. Hsu, A.~Mohamed,
  and E.~Dupoux, ``Speech resynthesis from discrete disentangled
  self-supervised representations,'' in \emph{arXiv}, 2021. [Online].
  Available: \url{https://arxiv.org/pdf/2104.00355.pdf}
\BIBentrySTDinterwordspacing

\bibitem{Kharitonov2021}
\BIBentryALTinterwordspacing
E.~Kharitonov, A.~Lee, A.~Polyak, Y.~Adi, J.~Copet, K.~Lakhotia, T.-A. Nguyen,
  M.~Rivi\`{e}re, A.~Mohamed, E.~Dupoux, and W.-N. Hsu, ``Text-free
  prosody-aware generative spoken language modeling,'' in \emph{arXiv}, 2021.
  [Online]. Available: \url{https://arxiv.org/pdf/2109.03264.pdf}
\BIBentrySTDinterwordspacing

\bibitem{Hasegawa-Johnson2017}
M.~Hasegawa-Johnson, A.~Black, L.~Ondel, O.~Scharenborg, and F.~Ciannella,
  ``Image2speech: Automatically generating audio descriptions of images,'' in
  \emph{ICNLSSP}, 2017, p. 1–5.

\bibitem{Wang2020-show-and-speak}
X.~Wang, S.~Feng, J.~Zhu, M.~Hasegawa-Johnson, and O.~Scharenborg, ``Show and
  speak: directly synthesize spoken description of images,'' in \emph{icassp},
  2021.

\bibitem{Hsu-etal-2021}
W.-N. Hsu, D.~Harwath, T.~Miller, C.~Song, and J.~Glass, ``Text-free
  image-to-speech synthesis using learned segmental units,'' in
  \emph{ACL-IJCNLP}, 2021, pp. 5284--5300.

\bibitem{Effendi2021}
J.~Effendi, S.~Sakti, and S.~Nakamura, ``End-to-end image-to-speech generation
  for untranscribed unknown languages,'' \emph{IEEE Access}, vol.~9, pp.
  55\,144--55\,154, 2021.

\bibitem{Ren2019-almost-unsup-tts}
Y.~Ren, X.~Tan, T.~Qin, S.~Zhao, Z.~Zhao, and T.-Y. Liu, ``Almost unsupervised
  text to speech and automatic speech recognition,'' in \emph{ICML}, 2019, pp.
  5410--5419.

\bibitem{Yeh2019-unsup-asr}
C.-K. Yeh, J.~Chen, C.~Yu, and D.~Yu, ``Unsupervised speech recognition via
  segmental empirical output distribution matching,'' in \emph{ICLR}, 2019.

\bibitem{Chen2019}
K.-Y. Chen, C.-P. Tsai, D.-R. Liu, H.-Y. Lee, and L.~shan Lee, ``Completely
  unsupervised speech recognition by a generative adversarial network
  harmonized with iteratively refined hidden {Markov} models,'' in
  \emph{Interspeech}, 2019.

\bibitem{Baevski2020-wav2vec2}
A.~Baevski, H.~Zhou, A.~Mohamed, and M.~Auli, ``wav2vec 2.0: A framework for
  self-supervised learning of speech representations,'' in \emph{Neural
  Information Processing Systems}, 2020.

\bibitem{Tachibana2018-DCTTS}
H.~Tachibana, K.~Uenoyama, and S.~Aihara, ``Efficiently trainable
  text-to-speech system based on deep convolutional networks with guided
  attention,'' in \emph{ICASSP}, 2018, pp. 4784--4788.

\bibitem{ljspeech17}
K.~Ito and L.~Johnson, ``The lj speech dataset,''
  \url{https://keithito.com/LJ-Speech-Dataset/}, 2017.

\bibitem{Panayotov15-LibriSpeech}
V.~Panayotov, G.~Chen, D.~Povey, and S.~Khudanpur, ``Librispeech: An {ASR}
  corpus based on public domain audio books,'' in \emph{ICASSP}, 2015, pp.
  5206--5210.

\bibitem{librilight}
J.~{Kahn}, M.~{Rivière}, W.~{Zheng}, E.~{Kharitonov}, Q.~{Xu}, P.~E.
  {Mazaré}, J.~{Karadayi}, V.~{Liptchinsky}, R.~{Collobert}, C.~{Fuegen},
  T.~{Likhomanenko}, G.~{Synnaeve}, A.~{Joulin}, A.~{Mohamed}, and E.~{Dupoux},
  ``Libri-light: A benchmark for asr with limited or no supervision,'' in
  \emph{ICASSP}, 2020, pp. 7669--7673.

\bibitem{g2pE2019}
K.~Park and J.~Kim, ``g2pe,'' \url{https://github.com/Kyubyong/g2p}, 2019.

\bibitem{Graves06-CTC}
A.~Graves, S.~Fern{\'{a}}ndez, F.~J. Gomez, and J.~Schmidhuber, ``Connectionist
  temporal classification: labelling unsegmented sequence data with recurrent
  neural networks,'' in \emph{ICML}, 2006, pp. 369--376.

\bibitem{Ott19-fairseq}
M.~Ott, S.~Edunov, A.~Baevski, A.~Fan, S.~Gross, N.~Ng, D.~Grangier, and
  M.~Auli, ``fairseq: A fast, extensible toolkit for sequence modeling,'' in
  \emph{Proceedings of NAACL-HLT 2019: Demonstrations}, 2019.

\bibitem{Povey11-kaldi}
D.~Povey, A.~Ghoshal, G.~Boulianne, L.~Burget, O.~Glembek, N.~Goel,
  M.~Hannemann, P.~Motlicek, Y.~Qian, P.~Schwarz, J.~Silovsky, G.~Stemmer, and
  K.~Vesely, ``The kaldi speech recognition toolkit,'' in \emph{ASRU}, 2011.

\bibitem{Hayashi20-espnet}
T.~Hayashi, R.~Yamamoto, K.~Inoue, T.~Yoshimura, S.~Watanabe, T.~Toda,
  K.~Takeda, Y.~Zhang, and X.~Tan, ``{ESPnet-TTS}: Unified, reproducible, and
  integratable open source end-to-end text-to-speech toolkit,'' in
  \emph{ICASSP}, 2020, pp. 7654--7658.

\bibitem{Kong20-hifigan}
J.~Kong, J.~Kim, and J.~Bae, ``{HiFi-GAN}: Generative adversarial networks for
  efficient and high fidelity speech synthesis,'' in \emph{Neural Information
  Processing Systems}, 2020.

\bibitem{park2019css10}
K.~Park and T.~Mulc, ``Css10: A collection of single speaker speech datasets
  for 10 languages,'' in \emph{Interspeech}, 2019.

\bibitem{hasegawa2020grapheme}
M.~Hasegawa-Johnson, L.~Rolston, C.~Goudeseune, G.-A. Levow, and K.~Kirchhoff,
  ``Grapheme-to-phoneme transduction for cross-language asr,'' in \emph{SLSP},
  2020, pp. 3--19.

\end{thebibliography}


\end{document}